# A Sigmoid-based car-following model to improve acceleration stability in traffic oscillation and following failure in free flow

Xingyu Chen, Haijian Bai

*Abstract*—This paper proposes an improved Intelligent driving model (Sigmoid-IDM) to address the problems of excessive acceleration in traffic oscillation and following failure in free flow. The Sigmoid-IDM uses a Sigmoid function to enhance the start-following characteristics, improve the output strategy of the spacing term, and stabilize the steady-state velocity in free flow. Moreover, the model's asymmetry is improved by means of introducing cautious following distance, driving caution factor, and segmentation function. The anti-interference ability of the Sigmoid-IDM is demonstrated by local stability and string stability analyses. The model parameters are calibrated based on the Hefei dataset and High D data for the start-up scene, stop-go scene, and free-flow scene. Compared with IDM, the Sigmoid-IDM significantly reduces errors and improves performance. In start-up and stop-go scenes, the RMSE for spacing is reduced by 46.71% and 30.48%, respectively. In free flow, the RMSE of velocity is reduced by 26.82%. Additionally, fuel consumption and comfort errors are reduced by 60.1% and 49.1%, respectively. Additionally, the simulated acceleration and deceleration of the Sigmoid-IDM are asymmetric in time and amplitude, better reflecting the following characteristics of human drivers. Circular road simulation and Simulink-Carsim Co-simulation of the Sigmoid-IDM are performed to verify the model's ability to reproduce traffic flow oscillations and the traceability of the model's planning trajectory.

## I. INTRODUCTION

Traffic oscillation, also known as stop-and-go traffic, refers to the phenomenon where vehicles experience repeated cycles of deceleration and acceleration in congested traffic conditions [1]-[4]. The car-following (CF) model serves as the fundamental basis for traffic flow studies and describes the longitudinal motion of vehicles. The primary goal of CF models is to simulate dynamic human following behavior. To achieve this goal, a calibrated following model can accurately regulate the movement of the following vehicle based on the real-time state of the preceding vehicle while meeting the driver's desired behavior. Additionally, the CF model must accurately represent realistic traffic flow states and reproduce observed traffic flow instabilities such as traffic oscillations.

In decades, various traditional CF models, such as Newell [5], OVM [6], GFM [7], and FVDM [8], that describe CF behavior have been extensively researched and reviewed by [9], [10]. However, they have limitations in setting parameters that lead to behavioral operations of acceleration and deceleration during driving that are inconsistent with driver expectations.

The time-continuous intelligent driving model (IDM) [11] ensures collision-free and self-organized driving. It generates realistic acceleration profiles and driver behavior in most single-lane traffic situations. Other modeling approaches lack IDM's well-defined property for each parameter through three common scenarios [12], i.e., free road acceleration from standstill, leading vehicle following, and slower or stopped vehicle approaching. For example, changing only $v_0$ and keeping other parameters constant simulates highway to city traffic transition, assuming active or defensive driving on both roads. Additionally, IDM's maximum acceleration and comfortable deceleration parameters avoid unrealistically high acceleration/deceleration absent in most earlier models [9]. However, IDM does not perform as perfectly in many special driving scenarios and often appears to contradict human driver behavior. Some studies modify the model from two perspectives: parameters and representation. This enables IDM to produce trajectories more consistent with driver expectations and explain traffic oscillation phenomena.

Several studies have focused on optimizing the IDM parameters for specific traffic flow conditions. The calibration of IDM parameters to suit different driving styles is crucial, as different drivers exhibit varied driving behaviors. [13] provides a comprehensive method for calibrating the IDM, which has been used in various datasets such as NGSIM [14], Shanghai [15], pNEUMA [16], and HighD [17],Waymo [18] .The IDM has shown accurate performance in both calibration and validation phases, as evidenced by low errors, which further validates its suitability for simulating traffic flow. [19] have simplified the IDM by performing parameter sensitivity analysis, which not only enhances the model's convergence velocity but also reduces the calibration's dimensionality. Similarly, [20]-[23] have conducted extensive research in this regard.[24] from the calibration trajectory indicated that IDM calibrates well in trajectory groups without stationary start situations, but calibration errors increase in trajectory groups with stationary start situations.

Additionally, studies have examined time-varying parameters to enhance the IDM's accuracy. The reason behind this is that such time-varying parameters can express the intra-driver heterogeneity. In recent years, various studies have attempted to optimize the IDM time-varying parameters to better simulate human driving behavior. IDMM (IDM with memory) [25] introduced an additional time-varying parameter to improve the calculation of the required time headway by considering the driver's memory effect in response to traffic conditions. [26] investigated the adaptive time headway as a function of local velocity variance in the IDM and proposed variance-driven time headways (VDT) to address the issue of T variations in long-time traffic congestion. Jiang [27] proposed 2D-IDM using the time-varying parameter T, taking into account the initial spacing of vehicles, which can change intentionally or unintentionally during the driving process. Furthermore,[28],[29] have designed and calibrated IDMs with time-varying parameters to better describe the different driving styles of drivers.

Calibrating IDM parameters using human driver trajectory data is a common approach to simulate human-like car-following behavior. While Punzo et al.[13] have made significant contributions to optimizing IDM parameters by improving parameter sensitivity, calibration algorithms, and objective functions, the simulated trajectories still differ from real trajectories. Intra-driver heterogeneity has an undeniable impact, and random parameterization is an effective method for reproducing traffic flow and heterogeneity. However, accurately identifying which parameter or parameters and how to identify the range of parameters in the random IDM can accurately represent traffic oscillations and intra-driver heterogeneity remains a challenge. Moreover, the use of machine learning for real-time calibration may also lead to the generation of a large amount of computation[29].

Other studies have modified the IDM expressions to fit specific scenarios, resulting in more accurate trajectory descriptions. Although several studies have attempted to optimize IDM expressions, limitations have been found in certain scenarios.[30] proposed, when entering a zone with a reduced velocity limit, the actual velocity may exceed the desired velocity, resulting in unrealistic deceleration. Similarly, during a vehicle lane change, the actual spacing may become much smaller than the desired spacing, causing the IDM to overreact in regaining the required spacing. Additionally, the influence of spacing may cause IDM to disperse to negative infinity at a given time [31]. On a macroscopic level, the fundamental diagram drawn by IDM does not exhibit a perfect triangular shape, as the IDM travels at a steady-state velocity in free flow, which is much lower than the desired velocity, resulting in an arc-shaped curve in free flow. [32]-[36] have shown that the density in free flow is related to the flow as a one-dimensional straight line of $Q=v_0\rho$. Some researchers suggest that setting the IDM acceleration exponent to tend towards positive infinity is the closest representation of the realistic fundamental diagram, but this parameter setting is not widely adopted. In the three-phase traffic theory proposed by Kerner [37], the steady-state of congested traffic is assumed to occupy a two-dimensional region in the flow density plane. Kerner states that there are three traffic phases: free flow, synchronous flow, and extensive moving congestion. In previous studies, IDM without the inclusion of stochastic parameters was unable to reproduce the inverse λ and synchronous congested traffic flow due to the limitation of the steady-state equilibrium equation. Although the IDMM and 2D-IDM reproduce the synchronous flow phenomenon by introducing dynamic T, the IDM steady-state equilibrium equation restriction also changes the IDM free flow profile.

Furthermore, several studies have pointed out some potential limitations of the IDM in accurately simulating human driver behavior in terms of acceleration and energy efficiency. For instance, [38] set the maximum acceleration parameter of the IDM to the theoretical maximum value observed in their experimental data, which required the driver to accelerate from a standstill to the maximum speed and then decelerate normally until the vehicle stopped again. However, this resulted in unrealistic and extreme acceleration and deceleration behaviors that deviated from human driving patterns. This casts doubt on the validity of the IDM in modeling acceleration behavior. Moreover, the IDM assumes that the maximum acceleration occurs at zero or low velocity, which does not reflect many real-world driving scenarios where drivers tend to reach a certain speed before accelerating to the maximum. This is confirmed by real vehicle data in [39], which indicates that the IDM's acceleration leads to a significant amount of energy waste in a stop-and-go environment that does not match human driving conditions. Additionally, many experimental studies of ACC models have shown that the IDM response is too slow and causes errors in the simulation of time headway compared to real data [10], [40].

Additionally, Wei [41] also found that driving behavior is asymmetric, meaning that vehicle acceleration and deceleration are not always equal in timing and magnitude. In reality, drivers often have a long delay and low acceleration when starting, but react fast and decelerate sharply to avoid potential collisions. The IDM is considered to take into account the driver's asymmetric behavior in its expressions[42]. However, in many simulation tests [43]-[45], IDM failed to simulate this asymmetric behavior (because no cruising process was observed) and was therefore unable to accurately simulate traffic oscillation caused by driver behavior. Furthermore, some drivers may prefer steady driving and not accelerate quickly, even with enough headway, to avoid future slowdowns and save fuel. But others may start faster, even with small headway, to drive more efficiently, though this may lower comfort and increase energy use. These driving style differences can cause stop-and-go phenomena, making it hard to model human driver behavior with the IDM.

The phenomenon of traffic oscillation, which is commonly described by the string instability of the CF model, has been extensively studied in [46]-[48]. However, it is important for the CF model to have the ability to restore stable traffic flow instead of perpetuating or worsening oscillations. Local instability is a significant issue that needs to be avoided to ensure the reasonableness of CF models [46]. To this end,[49] has proposed alternative models to improve the string stability of traffic flow, such as the effective cooperative CF model and the extended IDM with cooperation. Additionally, the modification of the deceleration term of IDM has been explored to improve its string stability [50]. [51] provides a framework for string stability of heterogeneous traffic flows and applies the framework to IDM to analyze the heterogeneity of model parameters between vehicles.

To solve the problems of unreasonable starting acceleration and unstable high-velocity in free flow caused by the IDM spacing strategy limitations, this paper proposes modifications from two perspectives: IDM parameters and IDM expression. In terms of IDM parameters, the addition of cautious following distance and cautious driving factor are proposed to better describe the cautiousness of different drivers at different driving phases. In terms of IDM expression, the Sigmoid function is introduced to adjust the spacing strategy, and the segmentation function is used to improve the asymmetry of IDM. The main contributions of this paper are as follows:

(1) The Sigmoid-IDM improves the response strength characteristics of the starting acceleration during the following start process by introducing the Sigmoid function, solving the problem of starting reversing generated by IDM. By also introducing the cautious following distance and cautious driving factor, the driving characteristics of drivers with different driving styles at start-up are more reasonably expressed, and the traffic oscillation phenomenon is reproduced more accurately.

(2) The Sigmoid-IDM more accurately simulates the asymmetric driving behavior of human drivers in terms of acceleration and deceleration in time and strength. This helps to more accurately represent real-world driving behaviors and traffic flow states.

(3) The Sigmoid-IDM solves the problem of unreasonable deceleration caused by spacing strategy limitations of IDM at high velocity, enabling the ability to drive more closely to the desired velocity steadily in free flow.

(4) Furthermore, we also propose a Sigmoid-IDM with time-varying cautious following distance considering the different caution levels of drivers at different initiation acceleration stages. And the model can reflect the heterogeneity within drivers and better simulate the fundamental diagram of traffic flow (Reproduce synchronous flow).

The remainder of this paper is organized as follows. Section II is definitions and limitations of IDM; Section III is model. Section IV is discussion on stability of Sigmoid-IDM and section V is numerical analysis. Finally, section VI concludes the paper and discussion.

## II. DEFINITIONS AND LIMITATIONS OF IDM

The Intelligent Driver Model (IDM) is a collision-free model proposed by Treiber based on the social force model in 2000. [15] indicates IDM simulates human CF behavior better compared to other models. The expression is as follows:

$$a_n(t) = a\left(1 - \left(\frac{v_n(t)}{v_0}\right)^\delta - \left(\frac{S^*}{S(t)}\right)^2\right) \quad (1)$$

$$S^* = s_0 + v_n(t)T + \frac{v_n(t)[v_n(t) - v_{n-1}(t)]}{2\sqrt{ab}} \quad (2)$$

$$a_{n1}(t) = a\left(1 - \left(\frac{v_n(t)}{v_0}\right)^\delta\right) \quad (3)$$

$$a_{n2}(t) = -a\left(\frac{S^*}{S(t)}\right)^2 \quad (4)$$

Where, $a_{n1}(t)$ is the free acceleration term; $a_{n2}(t)$ is the congestion deceleration term; $a$ is maximum acceleration; $b$ is desired deceleration; $v_0$ is desired velocity; $S(t)$ is spacing; $v_n(t)$ is following vehicle velocity; $v_{n-1}(t)$ is preceding vehicle velocity; $T$ is safe time headway; $s_0$ is jam distance; $\delta$ is acceleration exponent, and typical value is 4; $S^*$ is desired spacing.

When the traffic is in the state of free flow, the IDM uses the desired velocity as the stimulus to produce acceleration and deceleration, as shown in Eq. (3). When in congested flow, the IDM takes the desired spacing as the stimulus to produce acceleration and deceleration, as shown in Eq. (4). The velocity difference is taken as the stimulus for the change of desired spacing, so that the model can timely respond to the changes of the proceeding vehicle.

We will discuss the limitations of the IDM spacing strategy (without considering time-varying parameters), using data sets commonly used in existing studies to calibrate values of IDM parameters.

### A. IDM over-acceleration and negative velocity

When IDM is accelerated from parking, the change of initial spacing has an impact on the acceleration. When time $t = 0$, the starting acceleration can be expressed as Eq. (5).

$$a_n(0) = a\left(1 - \left(\frac{s_0}{S(0)}\right)^2\right) \quad (5)$$

The usual range of calibrated values of $a$ and $s_0$ is obtained from [29] as (0.5,4) and (0.1,3). Suppose the maximum acceleration $a = 3m/s^2$ and the jam distance $s_0 = 2m$. As shown in **Fig.1**, in scenarios where the preceding vehicle is stationary, the following driver is stimulated by the ratio of jam distance to the initial spacing, which may result in excessive starting acceleration exceeding 2.25 m/s² (yellow area). Although the initial spacing between the two vehicles may be only 4m at this point. This high starting acceleration can cause vehicle jerking and severely impact passenger comfort. Moreover, the IDM tends to initiate with high starting acceleration followed by abrupt braking, which is unsafe, especially in cases of small spacing. This is unacceptable for many cautious drivers.

Furthermore, when the initial spacing is less than the jam distance, the following vehicle's starting acceleration can become negative, leading to instances of vehicle reversing (red area). Such negative velocity in most micro-simulation frameworks, particularly under high-density simulation, is impossible and unrealistic. However, addressing the limitations of the IDM expression in these scenarios through calibration is challenging. The calibration process typically aims to minimize the global error across the trajectory, making it difficult to identify and rectify the local starting process's errors. Therefore, the inaccurate starting process of IDM is one of the factors contributing to the difficulty of calibrating driving trajectories in many traffic oscillation environments with high accuracy.

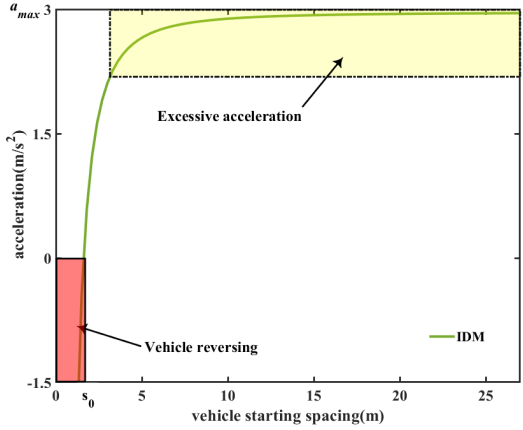

Fig 1. IDM starting spacing-acceleration curve from station (Two brief simulations about the limitations of IDM are in Appendix A)

*B Limitations of IDM High Velocity in the Free-Flow*

The High D dataset collected by Ika's team at the RWTH Aachen University [17] provides a more accurate dataset of highway vehicle trajectories. They tracked vehicles in aerial videos using U-Net. UAVs recorded traffic from above German freeways. The high D trajectories are mainly free-flow data, where the vehicle travels at a higher and more equilibrium velocity. [17] calibrates the IDM with a dataset where the average values of $v_0$ and $T$ are considered to be 30m/s and 1.5s.

The 3-vehicle platoon high-speed experiment was a CF experiment conducted by Jiang [52] on a high-speed 3-vehicle platoon on a circular highway in Hefei City. The platoon was not impeded by other vehicles and no vehicles switched in during the experiment because the traffic was free during the experiment time period (9:55 to 10:40 a.m.) and the experimental section was lanes 3 and 4. The leading cars in the experiment were asked to travel at different constant velocities at different time intervals.

In equilibrium traffic, IDM tends to keep an equilibrium spacing $S_e^{IDM}$ and equilibrium velocity $v_e^{IDM}$. The relationship between $S_e^{IDM}$ and $v_e^{IDM}$ is shown in Eq. (6).

$$S_e^{IDM} = \frac{s_0 + v_e^{IDM} T}{\sqrt{1 - \left(\frac{v_e^{IDM}}{v_0}\right)^\delta}} \quad (6)$$

Based on Eq. 6 we can plot the equilibrium state spacing-velocity diagram and the fundamental diagram in **Fig 2. (a)**.

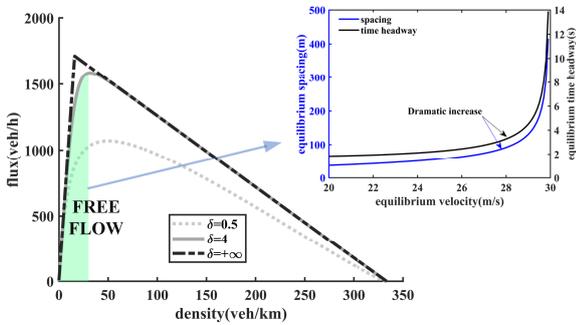

(a)The IDM equilibrium spacing- velocity and the fundamental diagram

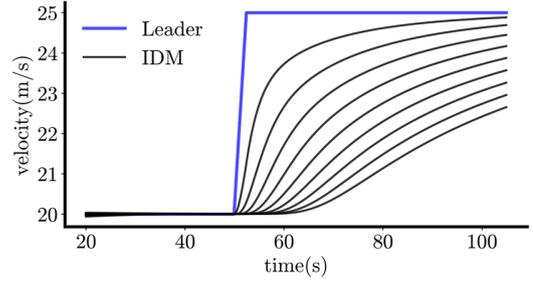

(b) Non-converging simulation of IDM driving at high velocity
Fig 2. IDM steady-state equilibrium analysis

In high-speed free flow, due to the free flow acceleration term Eq. (3) converging to 0, the IDM not only does not accelerate but also decelerates at $a_n(t) = -a[S^*/S(t)]^2$ when the spacing increases. This deceleration effect is even more pronounced in homogeneous traffic flow, although it enhances the string stability of the IDM. However, it is not appropriate to stabilize $a_n(t) = 0$ by keeping vehicles at high speeds only at $S^* \ll S(t)$. As shown in **Fig 2. (a)**, the IDM has much larger equilibrium spacing and time headway at high speeds. Therefore, in **Fig 2. (b)** a slight increase in the preceding vehicle's velocity causes a surge in the equilibrium spacing of the IDM, and the IDM needs to expand the spacing for a long time to maintain stability, which is unrealistic. This also makes the IDM response slow and causes errors between the velocity simulation and the real data.

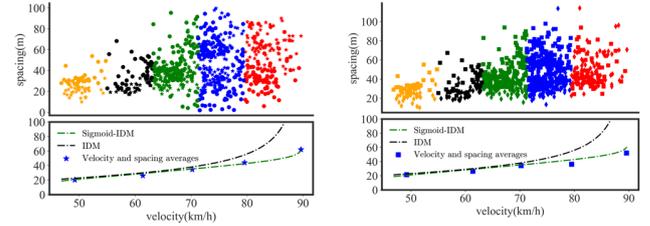

(a)Lane 3 platoon  (b)Lane 4 platoon
Fig 3. Spacing-velocity of the 3-vehicle platoon high-speed experiment

The total time for the 3-vehicle platoon high-speed experiment was 43 minutes. We collected the average values of trajectory velocity and spacing of the two following vehicles every five seconds and represented them with different shapes in **Fig 3**. when the vehicles are driving freely at high speed (red and blue areas), the spacing is unstable and varies more substantially. This indicates that within a certain range, the variation of the spacing does not have a significant effect on the vehicle velocity as long as the vehicle's free flow velocity is close to the desired velocity. Therefore, in the flow-density diagram, the slope of the free flow should be close to the constant desired velocity. However, as the vehicle approaches the desired velocity, the equilibrium spacing and time headway increase significantly due to the denominator of equation (6) tending to zero while the numerator tending to infinity. This prevents the vehicle from traveling in free flow at or near the desired velocity and makes the fundamental diagram arc-shaped (**Fig. 2 (a)**).

To verify this phenomenon, we fit the two models' the steady-state equations using the average spacing and the average velocity of the two lanes in **Fig 3**. As expected, the Sigmoid-IDM describes the steady-state spacing-velocity relationship well. In contrast, when the IDM steady-state velocity approaches the desired velocity, the data cannot be fitted well due to the surge in equilibrium spacing.

### III. MODEL

*A. Acceleration characteristics of start-to-follow*

The Sigmoid function is a common S-type curve in biology, also known as the S-type growth curve [53]-[55] as shown in **Fig.4**. The Sigmoid function expression is as follows: $f^\lambda(x) \in (0,1), x \in (-\infty, +\infty)$

$$f^\lambda(x) = \frac{1}{1+exp(-\lambda(x-x_0))} \quad (7)$$

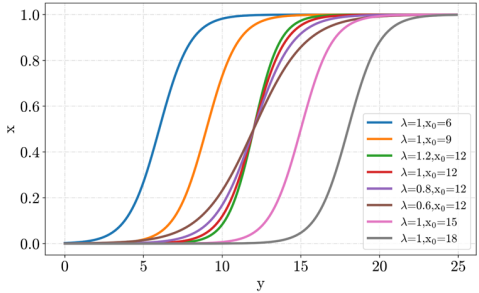

Fig 4. Sigmoid function with different parameters

The Hefei 25-vehicle platoon experimental data (Hefei data) used in this study is sourced from human-driven vehicle trajectories collected during 25 vehicles experiments on a three-lane suburban road in Hefei [52],[56]. High-precision differential GPS receivers were installed in each vehicle to record their position at 10 Hz. During each experiment, a leading vehicle followed a predetermined driving cycle, and the following vehicles started from a standstill and reached a certain velocity to achieve a steady-state driving condition. The purpose of using these data is to analyze the relationship between acceleration and spacing during vehicle start-up (orange area) and during stable driving (blue area) in **Fig 5(a)**.

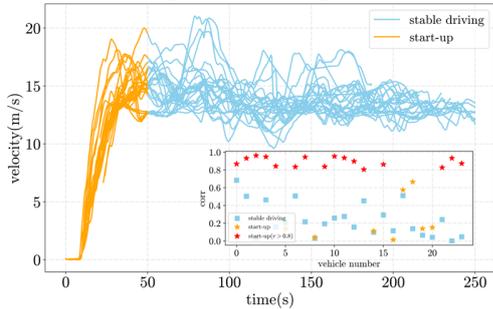

(a) Start-up trajectory extraction and correlation analysis

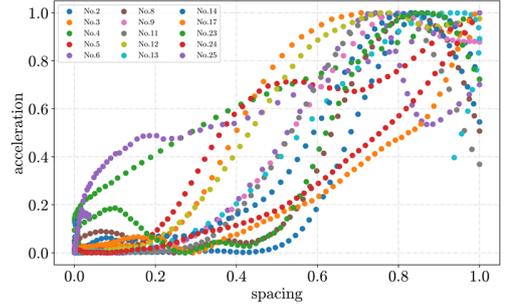

(b) Correlation trajectory acceleration-spacing normalization
Fig 5. Acceleration-spacing for the trajectories of the Hefei start-up process

We applied the method of [57],[58] for the reconstruction of the Hefei data trajectories, and analyzed the acceleration and spacing values of 24 following vehicles from trajectories after starting until stabilization. The results of the correlation analysis are shown in **Fig 5(a)**. The correlation between spacing and acceleration in the stable region is not significant (Pearson coefficients [59] are all less than 0.8), indicating that the driver responds less to change in spacing when the vehicle is driving stably at a velocity close to the desired velocity. However, this finding is not compatible with the setting of Eq. (6). Furthermore, during the start-up phase, the vehicles with a strong correlation between acceleration and spacing were screened by Pearson coefficient R>0.8. Sixteen vehicles met the requirement as shown in the **Fig 5(b)** and were used for further analysis. The spacing values of these vehicles were normalized to the acceleration values, and an "S"-like relationship （Not the quadratic function relationship of IDM）was found, indicating that the acceleration curve of human driving can be fitted by a Sigmoid function Eq. (7).

The curve fitting of vehicles' acceleration and spacing in start-up phase by Sigmoid function using MATLAB's curve fitting library, the parameters as well as the evaluation index of the effect are shown in **TABLE IV** in the appendix B, where the average value of $R^2$ of vehicles is 0.94, and the average value of RMSE is 0.096. The **Fig 6** shows the effect of starting curve fitting of the No.12 vehicle. Based on this similarity, we proposed the Sigmoid-IDM, which optimizes the vehicle's initial acceleration by means of a Sigmoid function with the aim of achieving a smooth start. And the cautious following distance and cautious driving factor are adapted to the different phases of vehicle driving.

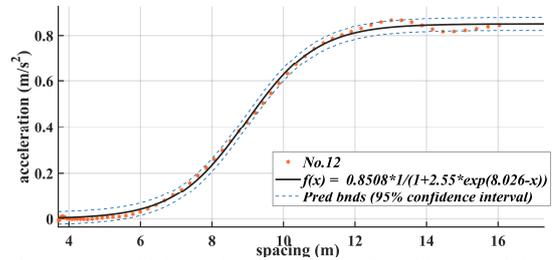

Fig 6. No.12 vehicle acceleration-spacing Sigmoid Curve Fitting

## B. Sigmoid-IDM

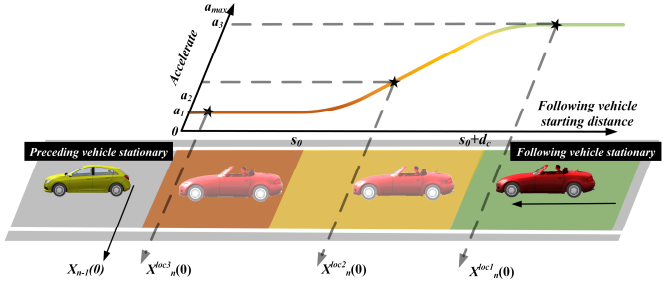

Fig.7. Acceleration indication of start-to-follow at different distance

Cautious following distance $d_c$ is the distance at which the driver will carefully choose small acceleration to follow the preceding vehicle. According to the Sigmoid curve, the start-to-follow acceleration of vehicles with different spacing is shown in **Fig.7**.

$S(0) \geq d_c + s_0$, drivers start with maximal acceleration to ensure efficiency.

$S(0) \leq s_0$, vehicle does not start.

$s_0 \leq S(0) \leq d_c + s_0$, the vehicle's initial acceleration increases in the "S" shape with the increase of $S(0)$.

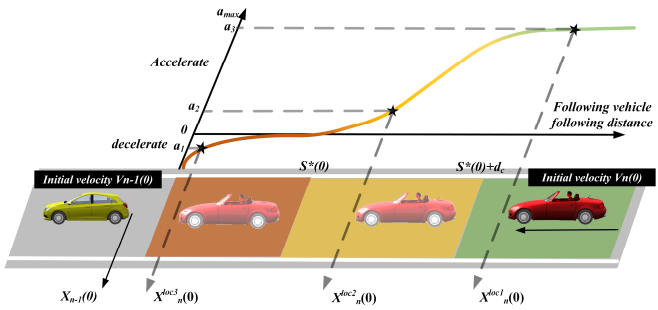

Fig.8. Acceleration Diagram of different spacing with an initial velocity

The acceleration diagram of the following vehicle with initial velocity at different spacing is shown in **Fig. 8**.

$S(0) \geq d_c + S^*(0)$, drivers start with maximal acceleration to ensure efficiency.

$S(0) \leq S^*(0)$, vehicle decelerates.

$S^*(0) < S(0) < d_c + S^*(0)$, the vehicle's initial acceleration increases in the "S" shape with the increase of $S(0)$.

In order to make the vehicle's initial acceleration meet the characteristics of S-curve, Sigmoid-IDM is built by adding Sigmoid function and cautious following distance into the spacing term of IDM. The model is expressed as following:

$$S^* = s_0 + v_n(t)T + \frac{v_n(t)[v_n(t) - v_{n-1}(t)]}{2\sqrt{ab}} \quad (8)$$

$$\begin{cases} if \quad S^* \geq S(t) > s_0 \\ a_n(t) = a\left(1 - \left(\frac{v_n(t)}{v_0}\right)^\delta - \left(\frac{S^*}{S(t)}\right)^2\right) \\ else \\ a_n(t) = a\left(1 - \left(\frac{v_n(t)}{v_0}\right)^\delta - \left[1 + \exp\left(\lambda(S(t) - S^* - d_c)\right)\right]^{-1}\right) \end{cases} \quad (9)$$
$$(10)$$

Where $d_c$ is cautious following distance; $\lambda$ is cautious driving factor, Sigmoid-IDM not only retains the advantages of IDM in deceleration scenarios, but also corrects the excessive acceleration of the original model by improving the acceleration term of IDM. At the same time, the acceleration and deceleration time of Sigmoid-IDM are asymmetrical, which is similar to the asymmetrical feature of human driving.

## C. Steady-state equilibrium equation

In equilibrium traffic, IDM tends to keep an equilibrium spacing $S_e^{S-IDM}$ and equilibrium velocity $v_e^{S-IDM}$. The relationship between $S_e^{S-IDM}$ and $v_e^{S-IDM}$ is shown in Eq.(11).

$$\begin{cases} S_e^{S-IDM} = \frac{1}{\lambda}\ln\left[\left(1 - \left(\frac{v_e^{S-IDM}}{v_0}\right)^\delta\right)^{-1} - 1\right] + s_0 + v_e^{S-IDM}T + d_c \\ if \quad s_0 + v_e^{S-IDM}T \geq S_e^{S-IDM} > s_0 \\ S_e^{S-IDM} = (s_0 + v_e^{S-IDM}T)\left(1 - \left(\frac{v_e^{S-IDM}}{v_0}\right)^\delta\right)^{-1/2} \end{cases} \quad (11)$$

Based on Eq. (11) we can plot the fundamental diagram in **Fig. 9(a)**.

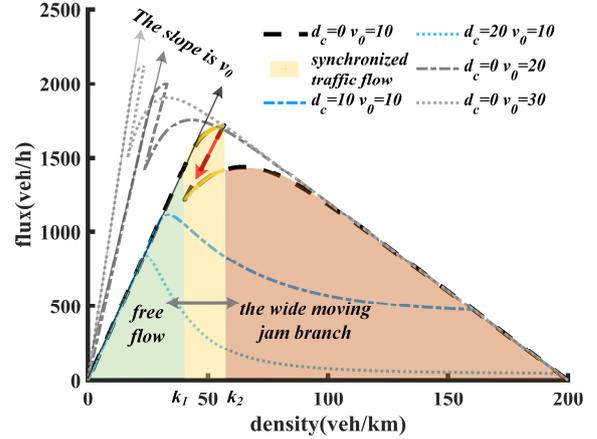

(a) The fundamental diagram of Sigmoid-IDM under different $d_c$ and $v_0$

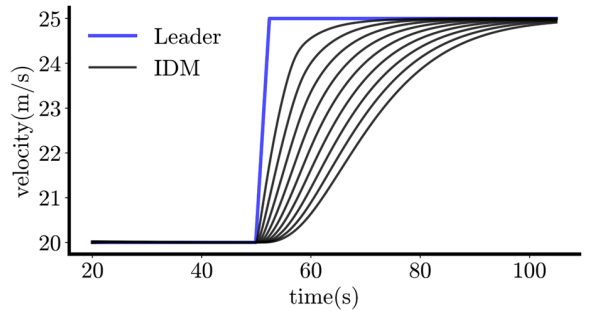

(b) Converging simulation of Sigmoid-IDM driving at high velocity
Fig.9. Sigmoid-IDM steady-state equilibrium analysis

Based on our simulation results, we observed that the IDM did not converge when it approached a stable velocity near the desired velocity in **Fig.2(b)**. In contrast, the Sigmoid-IDM was able to achieve quick convergence of fleet velocity when the velocity was close to the desired velocity in **Fig. 9(b)**. The fundamental diagram of the homogeneous flow was drawn using the steady-state equilibrium equation of the Sigmoid-IDM. The gray curves plotted in **Fig. 9(a)** for the examples $v_0 = 10, 20, 30$ indicated that the velocity of the free flow (green region) can maintain a steady desired velocity slope of $v_0$ i.e., $Q=v_0\rho$.

Moreover, as shown in **Fig. 9(a)**, we also found that the jam flow region (brown region) changes as the cautious following distance $d_c$ changes. Specifically, as $d_c$ increased from 10 to 20, the boundary line between the free flow and the wide density region (brown region) shifted to the left and the maximum capacity decreases, as shown by the blue line. Conversely, when $d_c$ decreased from 10 to 0, the maximum capacity dropped along the red arrow (black line). In this case, the traffic exhibited a hysteresis effect, where the free-flow and synchronous traffic flow coexisted in the state, and the flow-density curve took the inverse-λ form. A desirable property of the traffic flow model is its flexibility to represent various flow density shapes, from oblique parabolas to inverse -λ [60], which the Sigmoid-IDM is able to achieve by varying the value of $d_c$.

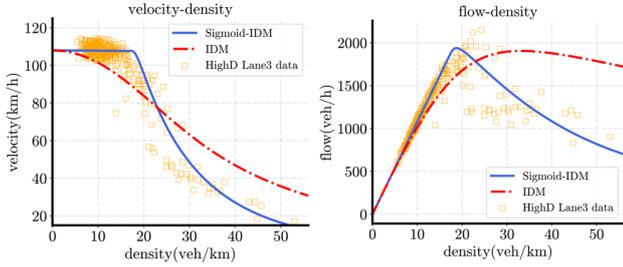

Fig.10. IDM and Sigmoid-IDM Fitted to High D lane3

The High D Lane 3 data is plotted in **Fig. 10**. the High-D data has a strong emphasis on the free-flow region, while the observations elsewhere are sparsely scattered. In addition, a distinctive feature of the flow-density subplot is the capacity peak, which clearly indicates an inverse-λ flow-density relationship. As expected, the Sigmoid-IDM is able to fit to such a shape and, therefore, can describe the free flow and capacity conditions well. In contrast, the IDM velocity-density curve is skewed and does not fit the flow-density relationship well for this data.

## IV. STABILITY

Stability is to describe the self-regulation ability of traffic flow with external disturbance. Local stability and string stability are usually used to verify the self-regulation ability of the model in single-vehicle following and multi-vehicle following scenarios respectively. The Sigmoid-IDM is functionalized as shown in Eq. (12). It is assumed that all vehicles in the stable traffic flow have the same spacing $s_t$ and velocity $v_t$ at time $t_s$, and the acceleration is 0, as shown in Eq. (13). And meanwhile, small changes in spacing, velocity or acceleration are regarded as small disturbances. The value of change in spacing $h_n(t)$ and the value of change in velocity $p_n(t)$ relative to the equilibrium state is calculated as shown in Eq. (14) and Eq. (15).

$$a_n(t) = f\left(S_n(t), v_n(t), \Delta v(t)\right) \quad (12)$$

$$a(t_s) = f(s_t, v_t, 0) = 0 \quad (13)$$

$$h_n(t) = S_n(t) - s_t \quad (14)$$

$$p_n(t) = v_n(t) - v_t \quad (15)$$

### A. Rationality of Following Behavior

In the following state, the vehicle should meet several rules [46] as following.
- When the spacing is larger, it should produce higher acceleration or smaller deceleration.
- When the velocity is high, the vehicle will not accelerate as much.
- The increase in relative velocity will cause the vehicle to accelerate and reduce braking.

If $s_0 + v_t T \geq s_t > s_0$:

$$f_s = \frac{\partial a}{\partial s}\bigg|t = 2a(s_0 + v_0 T)^2 / s_t^3 \quad (16)$$

$$f_v = \frac{\partial a}{\partial v}\bigg|t = -4av_t^3 / v_0^4 - (2a_t s_0 T + 2v_t T^2) / s_t^2 \quad (17)$$

$$f_{\Delta v} = \frac{\partial a}{\partial \Delta v}\bigg|t = -v_t(s_0 + v_t T)(a/b)^{1/2} / s_t^2 \quad (18)$$

$Else$:

$$f_s = \frac{\partial a}{\partial s}\bigg|t = a\lambda exp\left(\lambda(s_t - s_0 - v_t T - d_c)\right) \\ \times\left[1 + exp\left(\lambda(s_t - s_0 - v_t T - d_c)\right)\right]^{-2} \quad (19)$$

$$f_v = \frac{\partial a}{\partial v}\bigg|t = -4av_t^3 / v_0^4 - \lambda a exp(T(s_t - s_0 - v_t T - d_c)) \\ \times (1 + exp(\lambda(s_t - s_0 - v_t T - d_c)))^{-2} \quad (20)$$

$$f_{\Delta v} = \frac{\partial a}{\partial \Delta v}\bigg|t = (\lambda v_t(a/b)^{1/2} / 2)exp\left(\lambda(s_t - s_0 - v_t T - d_c)\right) \\ \times\left[1 + exp\left(\lambda(s_t - s_0 - v_t T - d_c)\right)\right]^{-2} \quad (21)$$

$$f_s > 0, f_v < 0, f_{\Delta v} > 0 \quad (22)$$

In typical scenario, a numerical test is performed as shown in **Fig.11**. When the stable velocity $v_t$ is 0-20 m/s and the stable spacing $s_t$ is 10-50 m, the solutions of the three partial derivatives of Sigmoid-IDM all meet the Eq. (22). Therefore, the following behavior of the Sigmoid-IDM is reasonable.
( $v_0 = 33.33 m/s$ , $b = 2m/s^2$ , $s_0 = 2m$ $a = 1.73 m/s^2$ , $\lambda=1$ , $T = 1s$ $d_c = 10m$ )

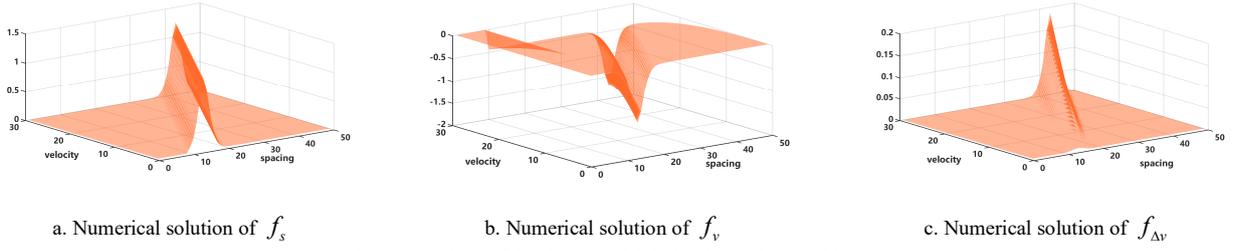

a. Numerical solution of $f_s$  b. Numerical solution of $f_v$  c. Numerical solution of $f_{\Delta v}$

Fig.11 Solutions of the three partial derivatives of Sigmoid-IDM

*B. Local stability*

Local stability refers to whether the state changes of the following vehicle converge when the preceding vehicle is disturbed. When a small disturbance is loaded on the preceding vehicle, the stable spacing and velocity difference will change. If the model is local stability, the following vehicle will eventually maintain the stable spacing or same velocity with the preceding vehicle through acceleration and deceleration. The Taylor expansion of Eq. (12) yields Eq. (23). Then, Eq. (14) and Eq. (15) are substituted into Eq. (23) to be Eq. (24).

$$a_n(t) \approx f(s_t, v_t, 0) + \frac{\partial a}{\partial s}\big|_t (S_n(t) - s_t) + \frac{\partial a}{\partial v}\big|_t (v_n(t) - v_t) + \frac{\partial a}{\partial \Delta v}\big|_t (\Delta v(t) - 0) \quad (23)$$

When the following vehicle is stable at $t_s$:

$$a_n(t_s) = f_s h_n(t_s) + (f_v - f_{\Delta v}) p_n(t_s) + f_{\Delta v} p_{n-1}(t_s) \quad (24)$$

When the following vehicle is in a stable state, the preceding vehicle should also be in a stable state. Thus, $p_{n-1}(t_s) = 0$, the Eq. (24) is transformed into the Eq. (25)

$$\frac{dh_n(t)}{dt} = -p_n(t)$$
$$\frac{dp_n(t)}{dt} = \left(f_s h_n(t) + (f_v - f_{\Delta v}) p_n(t)\right)_{t_s} = 0 \quad (25)$$

From [48], the local stability of the CF model can be guaranteed when the eigenvalue $\gamma$ of the matrix $J_e = \begin{bmatrix} 0 & -1 \\ f_s & f_v - f_{\Delta v} \end{bmatrix}$ has two negative real parts. The calculation of the eigenvalue $\gamma$ is $|J_e - \gamma I| = 0$ ($I$ is the unit diagonal matrix). The eigenvalue equation can be translated as:

$$\gamma^2 - (f_v - f_{\Delta v})\gamma + f_s = 0$$
$$\gamma_\pm = (f_v - f_{\Delta v} \pm \sqrt{(f_v - f_{\Delta v})^2 - 4f_s})/2 \quad (26)$$

The characteristic equation-based method aims to evaluate the perturbation growth rate from the mathematical solution, which can be used to verify the local stability of the model. When the real parts of the two solutions ($\gamma_\pm$) are both negative, the model is locally stable. According to Eq. (20), if the following behavior is reasonable, then $f_v - f_{\Delta v} < 0$. Therefore, the negative real part solution of Eq. (24) must exist and the model is local stability.

*C. String stability*

String stability is a property of the backward propagation of state changes from the leading vehicle in a following fleet. In a stable fleet, if a Fourier disturbance affects the velocity of the leading car, the transfer state of the disturbance in the fleet can be analyzed to determine whether the transfer state is stable. According to Fourier theory, periodic signals can be decomposed into the sum of periodic sine and cosine waves only. We assume the disturbance of the leading car is stable oscillation $h_0(t) = exp(iwt)$. Let $G(s)$ ($s = iw$) be the transfer function, then the oscillation of the $n^{th}$ following vehicle in the convoy is Eq. (27). Then, Eq. (14), Eq. (15) and Eq. (27) are substituted into Eq. (24) to be Eq. (28).

$$h_n(t) = (G(iw))^n exp(iwt) \quad (27)$$

$$G(iw) = \frac{f_s + iw f_{\Delta v}}{-w^2 + f_s - wi(f_v - f_{\Delta v})} \quad (28)$$

The string stability of the model is proved when $G(iw) < 1$. Therefore the string stability criterion is $w^2 f_{\Delta v}^2 + f_s^2 < (f_s - w^2)^2 + w^2(f_v - f_{\Delta v})^2$. When $w \to 0$, the lower the frequency is, the stronger the constraint on stability is. Thus satisfying Eq. (29) means that the model is string stable.

$$\frac{1}{2} - \frac{f_{\Delta v}}{f_v} - \frac{f_s}{(f_v)^2} > 0 \quad (29)$$

The maximum velocity is set to $v_0 = 78 km/h$, and the equilibrium velocity is set to 25 km/h to highlight the influence of the parameters on the string stability of the model [51]. The influence parameters of driver aggressiveness of the Sigmoid-IDM are represented by $\lambda$ and $d_c$. In [11], $a$ and $T$ were selected as variables due to their significant impact on the string stability of the IDM, and the string stability behavior criterion of the Sigmoid-IDM was illustrated using Eq. (29). The string stability of the Sigmoid-IDM under various values of $\lambda$ and $d_c$ is shown in **Fig. 12**, and the string stability behavior of the Sigmoid-IDM is depicted within its parameter variable range. The unstable region is depicted in pink, whereas the stable region is represented in white.

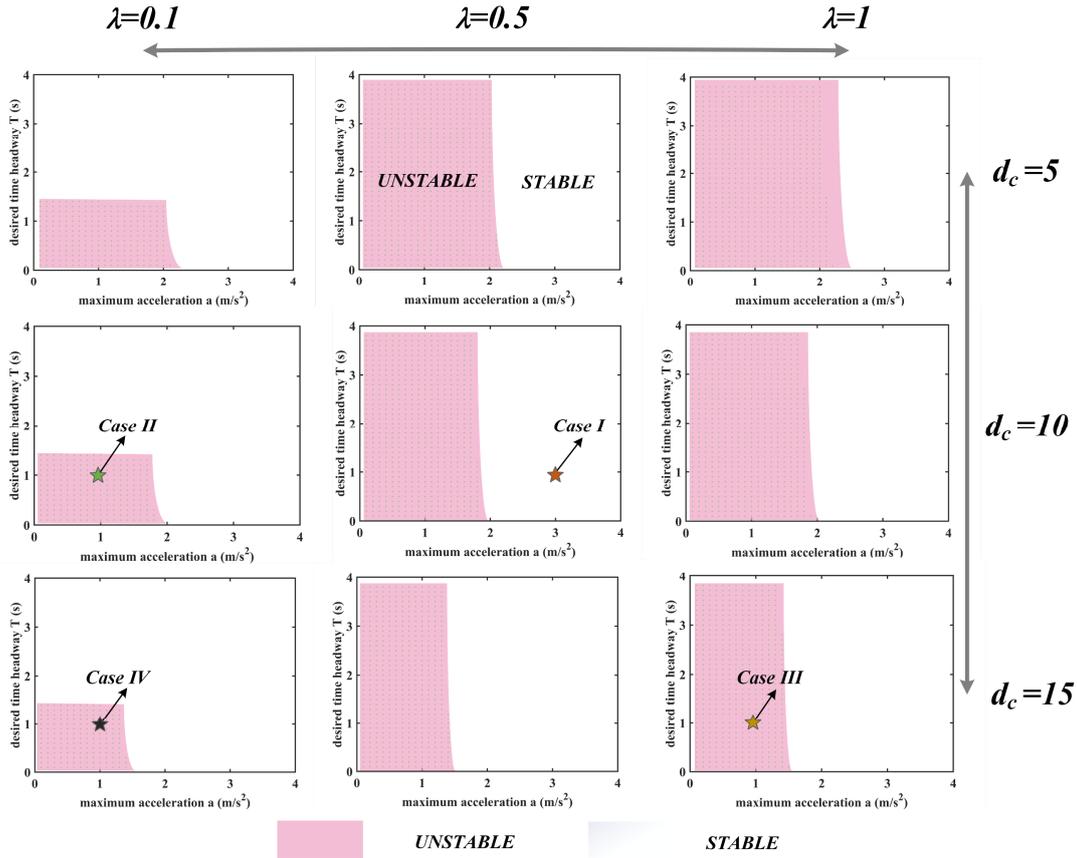

Fig.12 String stability diagrams of the Sigmoid-IDM for heterogeneous vehicles

The Sigmoid-IDM can simulate traffic oscillation scenarios by adjusting $\lambda$ and $d_c$, capturing excessive acceleration and deceleration to produce acceleration/deceleration string instability. To validate the string stability criterion and analyze the effect of $\lambda$ and $d_c$ on the acceleration and deceleration of the string in an unstable state, a simulation with 11-vehicle platoon is conducted. We observe whether any perturbation in velocity will amplify or not. The selection of parameters for each case is shown in **Fig. 12**. And the leading vehicle's velocity is set to a sinusoidal velocity profile. The resulting velocity-time diagram of the following vehicles is plotted in **Fig. 13**.

- **Case I** is the simulation of the stable region for $d_c = 10, \lambda = 0.5$. It can be found that the model is string stable at this time.
- **Case II** is the simulation of the unstable region at $d_c = 10, \lambda = 0.1$. In this case the deceleration is string stable and the acceleration is string unstable. The velocity perturbation is amplified during acceleration and is more responsive during the start-up phase.
- **Case III** is simulated for $d_c = 15, \lambda = 1$ in the unstable region. In this case the acceleration is string stable and the deceleration is string unstable. The velocity perturbation is amplified during deceleration and there is a significant delay in waiting during the starting phase of the vehicle, which is often similar to that of many cautious drivers in reality and illustrates the role of cautious following distance introduction.
- **Case IV** is simulated for $d_c = 15, \lambda = 0.1$ in the unstable region. In this case the acceleration and deceleration are string unstable and over-acceleration and deceleration can occur. Therefore, the driving characteristics of different drivers, i.e., acceleration and deceleration characteristics, can be simulated by reasonably calibrating the values of $\lambda$ and $d_c$ to present asymmetric driving behavior.

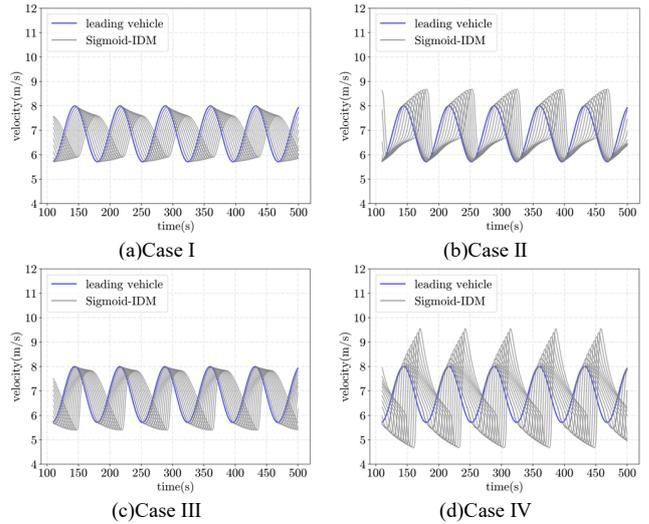

Fig. 13 Simulation of eleven vehicles in different parameter cases.

## V. NUMERICAL ANALYSIS

In this section, we calibrate the Sigmoid-IDM and IDM models, using the above-mentioned Hefei dataset and the High D dataset to obtain the necessary parameter values. Following this, we evaluate the accuracy of the models by simulating different scenarios and comparing the results with experimental data using error measures. Subsequently, we analyze comfort and fuel consumption data, which indicates that Sigmoid-IDM performed better than IDM in managing starting acceleration and better represented human driving behavior. We also examine the simulated spacing-velocity relationship to confirm the asymmetry of Sigmoid-IDM. Then, we introduce random parameters into the Sigmoid-IDM and study the resulting traffic flow oscillation phenomenon, ultimately obtaining the fundamental diagram through the Circular Road simulation. Finally, the traceability of the Sigmoid-IDM trajectory was verified by Simulink-Carsim Co-simulation.

*A. Calibration*

To investigate the performance differences between human driving, IDM and Sigmoid-IDM in start-stop scenarios, stop-and-go scenarios and free-flow scenarios, the models were calibrated using 72 sets of experimental data, 48 vehicles from the Hefei data (24 for start-stop scenarios and 24 for stop-and-go scenarios) and 24 vehicles from the high D data (free-flow scenarios). Twelve of the follower trajectories in each scenario are used for calibration and the rest for validation. The calibration method [61] uses a genetic algorithm (GA) with spacing chosen as the performance metric and a population size equal to 100 allowing the algorithm to possibly converge to a global optimal solution with a maximum number of 500 iterations. The objective function is Eq. (30) for minimizing spacing errors and the error formula is Eq. (31) for checking the calibration effect. The iterative calibration results are shown in **TABLE I**. In addition, we performed 20 iterations of each calibration experiment to verify that the algorithm converged to the same solution for each calibration for both the parameter values and the objective function values (the standard deviation recorded between solutions at convergence was equal to $10^{-7}$).

$$U(s) = \frac{\sqrt{\frac{1}{T}\sum_{t=1}^{T}\left[s_{obs}(t) - s_{sim}(t)\right]^2}}{\sqrt{\frac{1}{T}\sum_{t=1}^{T}\left[s_{obs}(t)\right]^2} + \sqrt{\frac{1}{T}\sum_{t=1}^{T}\left[s_{sim}(t)\right]^2}} \quad (30)$$

$$RMSE_{(s,v,a)} = \sqrt{\frac{1}{T}\sum_{t=1}^{T}\left[(s,v,a)_{obs}(t) - (s,v,a)_{sim}(t)\right]^2} \quad (31)$$

*B. Start-up scenes*

The simulation results in **TABLE II** show that in the start scenes, the average RMSE of the 12 following vehicle trajectories simulated by Sigmoid-IDM is significantly smaller than that of IDM when the preceding vehicle is also in the start scenario. The prediction accuracy of spacing, velocity, and acceleration of Sigmoid-IDM is also significantly better, improving by 30.48%, 28.84%, and 27.27%, respectively, compared to IDM. This paper argues that the simulation error difference between the two models is mainly reflected in the vehicle start-up phase. We also analyze the variability between the model and the actual data in terms of both fuel consumption and comfort.

**TABLE I**
CALIBRATION VALUES OF THE MODELS IN DIFFERENT SCENES

| Scenes | Model | $\bar{a}$(0.1-6) | $\bar{b}$(0.1-6) | $\bar{v_0}$(10-40) | $\bar{T}$(0.1-4) | $\bar{s_0}$(0.1-6) | $\bar{\lambda}$(0-2) | $\bar{d_c}$(0.1-20) | $\delta$ | $U(s)$ |
|---|---|---|---|---|---|---|---|---|---|---|
| Start-up (Hefei) | IDM | 2.8 | 4.9 | 22.1 | 2.6 | 1.3 | \ | \ | 4 | 0.23 |
| | Sigmoid-IDM | 4.9 | 5.1 | 23.8 | 2.4 | 1.1 | 0.9 | 11.9 | 4 | 0.13 |
| Stop-and-go (Hefei) | IDM | 2.4 | 3.7 | 21.1 | 3.1 | 0.7 | \ | \ | 4 | 0.15 |
| | Sigmoid-IDM | 4.3 | 5.3 | 23.8 | 2.8 | 0.71 | 0.8 | 10.7 | 4 | 0.08 |
| Free-flow (High D) | IDM | 2.7 | 4.8 | 39.9 | 1.5 | 1.73 | \ | \ | 4 | 0.05 |
| | Sigmoid-IDM | 4.8 | 5.9 | 38.2 | 1.3 | 1.65 | 0.8 | 17.4 | 4 | 0.03 |

**TABLE II**
ERROR VALUES OF THE MODELS IN DIFFERENT SCENES

| Scenes | Model | $\overline{RMSE_s}$ | diff | $\overline{RMSE_v}$ | diff | $\overline{RMSE_a}$ | diff |
|---|---|---|---|---|---|---|---|
| Start-up (Hefei) | IDM | 6.13 | 30.48% | 0.52 | 28.84% | 0.22 | 27.27% |
| | Sigmoid-IDM | 4.26 | | 0.37 | | 0.16 | |
| Stop-and-go (Hefei) | IDM | 4.73 | 46.71% | 0.39 | 41.02% | 0.17 | 41.17% |
| | Sigmoid-IDM | 2.52 | | 0.23 | | 0.10 | |
| Free-flow (High D) | IDM | 3.41 | 2.05% | 0.41 | 26.82% | 0.49 | 4.10% |
| | Sigmoid-IDM | 3.34 | | 0.30 | | 0.47 | |

To illustrate the advantages of Sigmoid-IDM over IDM, an example (No.13) is provided in **Fig. 14**, which shows the acceleration, velocity and spacing curve of the actual vehicle during start-up. It is observed that the driver keeps an observation phase before starting, during which the spacing remains stable, and then chooses to start when the current distance reaches a certain value. However, IDM does not simulate this effect and starts directly. Moreover, the acceleration profile of the driver during start-up increases slowly with increasing velocity, which is consistent with [39] experimental data. In contrast, the IDM starts directly at maximum acceleration and decelerates immediately after reaching a certain velocity in a short time, leading to energy waste and a poor comfort level.

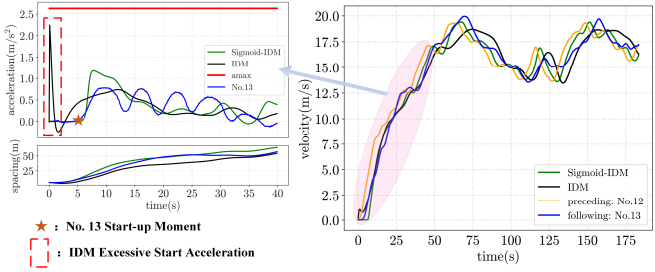

Fig. 14. IDM and Sigmoid-IDM trajectory of simulation No.13

The Sigmoid-IDM, on the other hand, exhibits a high human-like effect throughout the simulation, and the fuel consumption (**Fig. 15**) and comfort (**Fig. 16**) are more similar to the real trajectory data. Sigmoid-IDM reduces fuel consumption and comfort errors by 60.1% and 49.1% compared to IDM. Although it is not possible to show the simulation results for all starting vehicle trajectories, the presented example and the overall analysis suggest that Sigmoid-IDM has significant advantages over IDM in accurately simulating the start-up phase of vehicles.

Comfort is an important criterion to evaluate the intensity of driving behavior [62]. Comfort is the subjective feeling of passengers, mainly affected by the impact force. The jerk (derivative of acceleration) is usually used to reflect the impact force. The smaller the jerk is, the higher the comfort level of passengers will be. Conversely, the worse the comfort level of passengers will be.

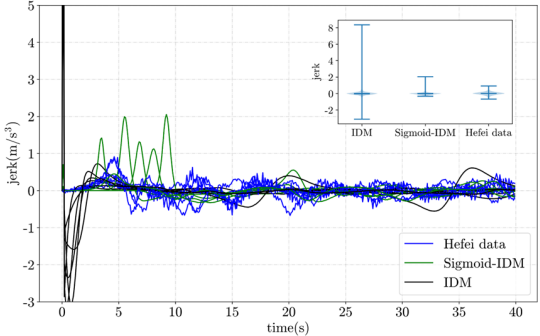

Fig. 15. Real-time comfort (jerk) of IDM and Sigmoid-IDM's simulation

The fuel consumption of a vehicle is closely related to the instantaneous velocity and acceleration of the vehicle. The VT-Micro model proposed by [63] is widely used in traffic flow microsimulation and is calculated as follows.

$$\ln(MOE_n) = \sum_{i=0}^{3}\sum_{j=0}^{3} k_{i,j}(v_n)^i(a_n)^j \tag{32}$$

where $MOE_n$ is the fuel consumption factor for the $n^{th}$ vehicle; $i = 0, 1, 2, 3$ is the index of the speed; $j = 0, 1, 2, 3$ is the index of the acceleration; and $k_{i,j}$ is the regression coefficient when the velocity index is $i$ and the acceleration index is $j$. With the calibrated $k_{i,j}$ values [64], the fuel consumption of the vehicle $n$ at each moment can be calculated. The violin plot (**Fig. 16**) counts the total value of fuel consumption in the start-up phase for each vehicle in the Hefei data, and the total value of fuel consumption in the vehicle simulation in the start-up phase of IDM and Sigmoid-IDM is used as a comparison.

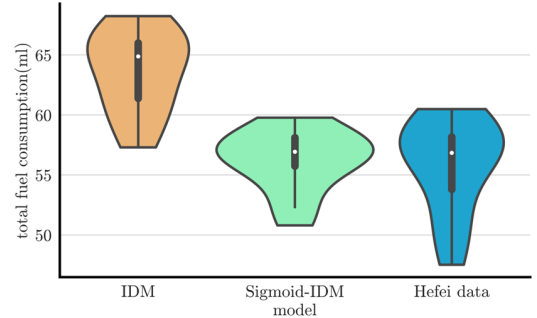

Fig. 16. IDM and Sigmoid-IDM simulation of the total fuel consumption distribution per vehicle

### C. stop-and-go scenes

Stop-and-go traffic is a prevalent and complex phenomenon in traffic congestion. The CF model replicates human driver behavior, including asymmetry and traffic oscillations. Previous analyses have shown the proposed Sigmoid-IDM's ability to simulate the vehicle start-up phase. To further validate its performance in stop-and-go traffic, we designed a simulation experiment with 24 vehicles in a single lane. We assigned the lead vehicle's velocity as the blue line in **Fig. 17**, based on the observed data from the Hefei dataset. The following vehicles used the calibrated Sigmoid-IDM-based CF model. **Fig. 17** shows the simulated velocity of the vehicles. The lead vehicle's perturbation increases as it spreads upstream, while the following vehicles experience stronger velocity changes, creating stop-and-go traffic in the platoon. This demonstrates the Sigmoid-IDM's ability to capture real traffic features.

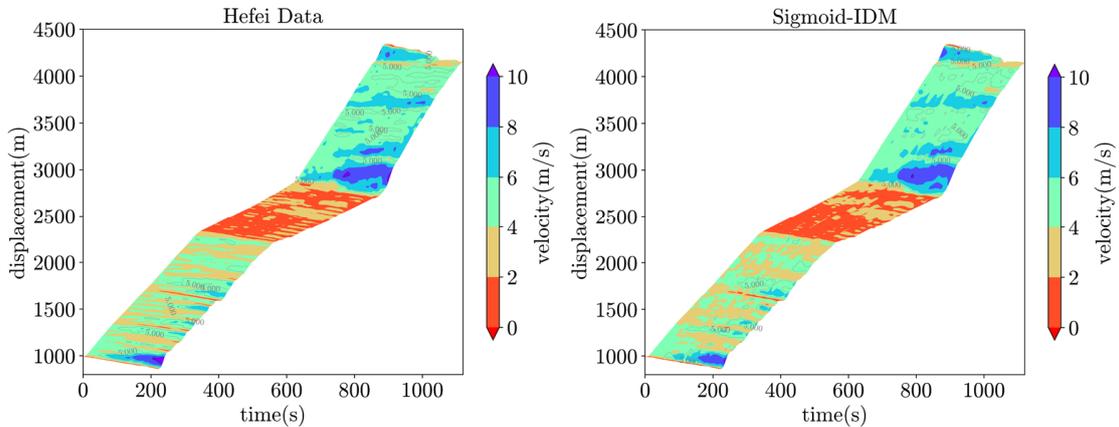

a. Time-space diagrams of Hefei data and Sigmoid-IDM in stop-and-go traffic

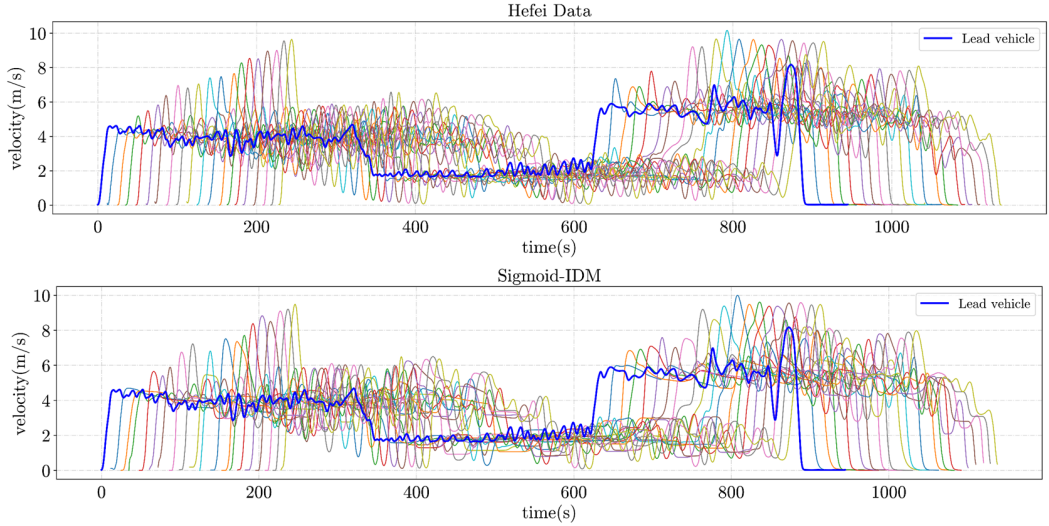

b. Time-velocity diagrams of Hefei data and Sigmoid-IDM in stop-and-go traffic
Fig. 17 Sigmoid-IDM stop-and-go scene simulation

The CF model's asymmetry is essential for understanding traffic hysteresis and stop-and-go phenomena [43]-[45]. In stop-and-go conditions, vehicles undergo various phases: acceleration, deceleration, and coasting. During acceleration, vehicles increase their velocity slowly and maintain a wide spacing to avoid sudden stops by the preceding vehicle. During deceleration, vehicles reduce their velocity and use as little spacing or delayed braking as possible to prepare for the preceding vehicle's stop. During coasting, vehicles keep a constant velocity between the acceleration and deceleration curves. The spacing changes depending on the preceding vehicle's acceleration or deceleration. The coasting phases indicate asymmetric behavior in traffic, as longer coasting periods imply larger distances between acceleration and deceleration curves. We examine the Sigmoid-IDM's asymmetric behavior by simulating spacing-velocity plots for the two models for No. 3 and No. 10's stop-and-go state in **Fig 18**. The simulation does not observe any coasting phase (black) in the IDM, and the acceleration (blue) and deceleration (red) curves are not separated. In contrast, the Sigmoid-IDM shows a sufficient coasting phase between the acceleration and deceleration phases. The Sigmoid-IDM reflects asymmetry better than the IDM and aligns more with human drivers' style in stop-and-go conditions.

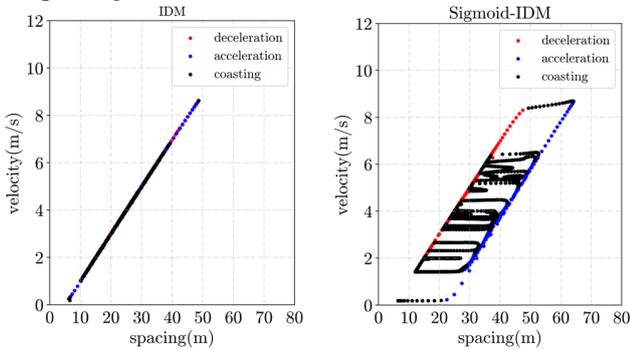

（a）No. 10

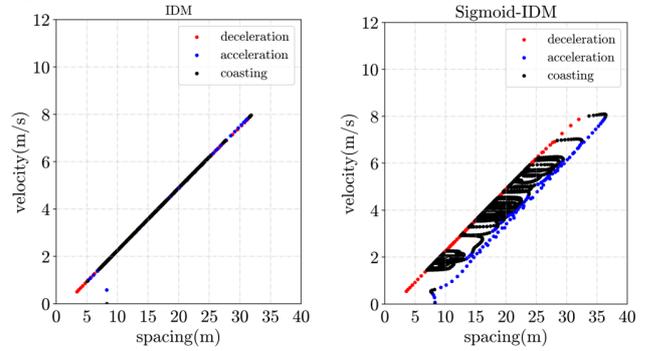

（b）No. 3

Fig. 18. Vehicle's spacing-velocity diagram of IDM and Sigmoid-IDM

### D. Circular Road

In this section, two simulations of Sigmoid-IDM are performed on the circular road. Simulation 1 is performed in order to verify that the proposed cautious following distance $d_c = 0$ can reproduce the synchronous flow. Simulation 2 is performed to explore the correlation between the time-varying $d_c$ on the driver's cautious driving and the effect on the flow-density relationship in traffic oscillations.

**Simulation 1: Simulation of simultaneous traffic flow**

We simulate the traffic flow on a circular road to investigate whether Sigmoid-IDM can simulate the synchronous traffic flow to reproduce the fundamental diagram drawn by the steady-state equilibrium equation in Section III when the cautious following distance $d_c = 0$. In the simulation, the parameters are set to the same parameters as in Section III C and additionally given as *a=1.5, b=3*. The following two initial configurations are used in the simulation: (1) all vehicles are homogeneously distributed on the road; (2) all vehicles are jam distributed.

**Fig. 19.** shows the flow-density diagram of the Sigmoid-IDM. In the branches with density less than $k_1$, there is only free flow on the road. There are two branches in the density region $k_1 < k < k_2$. The upper branch exhibits synchronous flow, which starts from an initial homogeneous distribution; the lower branch is a coexisting state of free flow and wide moving jams,

which comes from an initial jam distribution. When the density is greater than $k_2$, the synchronous flow is unstable and finally a wide moving jam occurs.

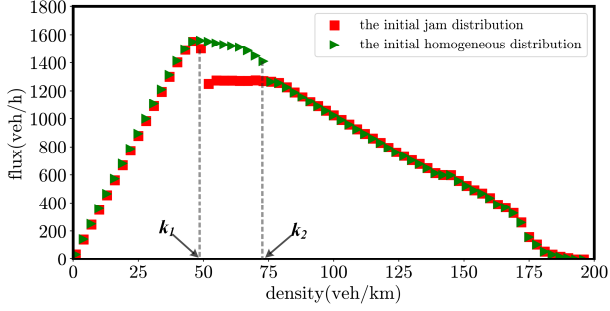

Fig. 19. Flow-density diagrams of Sigmoid-IDM at $d_c$=0.

**Simulation 2: Random Parameter Traffic Flow Simulation**

In this section, we propose a variable parameter form for the proposed Sigmoid-IDM, where $d_c$ is time-varying cautious following distance. by varying the value of $d_c$ for different driving states to explore the effect of having the intra-driver heterogeneity under Sigmoid-IDM on traffic oscillations. To ensure heterogeneity among drivers, in the simulation we use the calibrated values of the 20 vehicles in the stop-and-go scenario in section A above, and the $d_c$ changes its value in each simulation step $\Delta t = 0.1$ s. The time-varying expression of is shown in Eq. (33).

$$d_c(t+\Delta t) = \begin{cases} d_c(t) + rd & \text{with probability } p \\ d_c(t) & \text{otherwise} \end{cases} \quad (33)$$

Where, $d$ is a parameter indicating the range of cautious following distance variation; $r$ is an independent random number between -1 and 1; $p$ is the parameter change probability.

We simulated traffic flow on a 600m circular road, with each vehicle set to a length of 5m. The fleet was initialized with an initial velocity of 5m/s and an initial spacing of 10m. The total simulation time is 3600s, and the simulation step size is 0.1s. We set $p=0.25, d = 5$ and then analyzed three cases for the value of $r$: first, $r$ was set as a random value from -1 to 0 (*max* ($d_c$, 0)); second, $r$ was set to 0; and third, $r$ was set as a random value from 0 to 1. For each case, we analyzed the effect of changes in $d_c$ on traffic flow oscillation, and space-time diagrams were generated (**Fig.20**). Next, we plotted flow-density diagrams for the three cases with different densities (ranging from 0 to 200veh/km) to verify the analysis of the steady-state equilibrium equation discussed in Section III C.

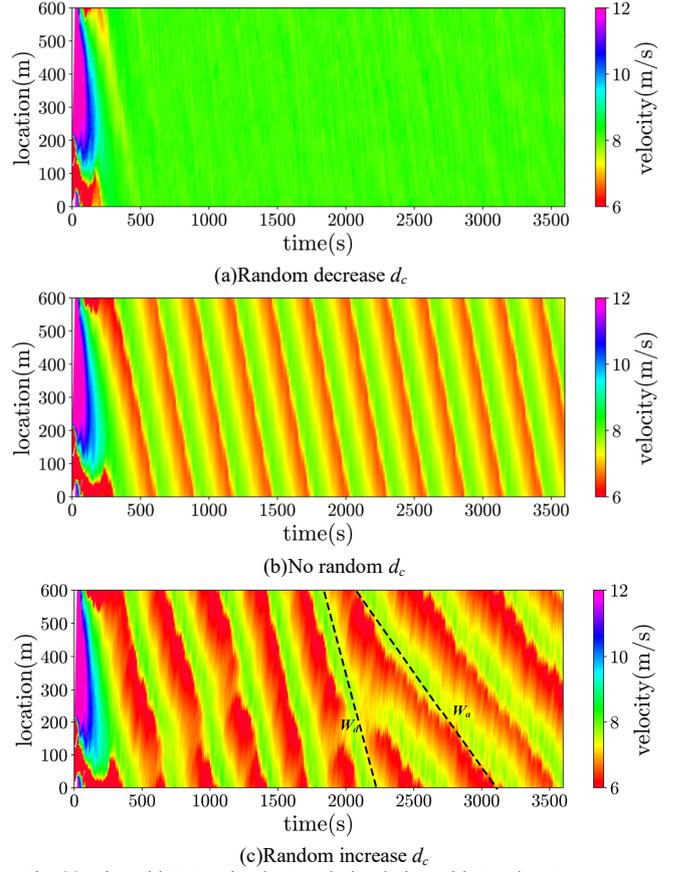

Fig. 20. Sigmoid-IDM Circular Road Simulation with Random Parameters

The **Fig.20(b)** shows that the 20 vehicles in the traffic flow oscillation data calibrated with the Sigmoid-IDM simulation, without adding the time-varying random cautious following distance, display a stop-and-go state for the fleet driving in the circular road. However, the average velocity and oscillation frequency of the traffic flow significantly change after incorporating the random $d_c$. When the random $d_c$ decreases as shown in **Fig. 20(a)**, the driver's aggressiveness increases, resulting in faster vehicle acceleration responses and a reduction in traffic flow oscillation. Conversely, as depicted in **Fig. 20(b)**, increasing the random $d_c$ results in a higher degree of driver caution and a slower fleet acceleration response. The random variation of drivers leads to different levels of driver caution at different moments, which is manifested in different acceleration responses due to the asymmetric of Sigmoid-IDM. At 2500s, a "secondary disturbance" phenomenon is observed in the oscillation region due to vehicle over-response or under-response, resulting in the acceleration wave $W_a$ not being equal to the deceleration wave $W_b$. The wave speed of the traffic flow oscillation is then maintained at the magnitude of $W_a$ in the **Fig.20(c)** afterwards. The fact that $W_b$ is smaller than $W_a$ is due to the increase in $d_c$ and the relative slower propagation of traffic congestion for cautious drivers.

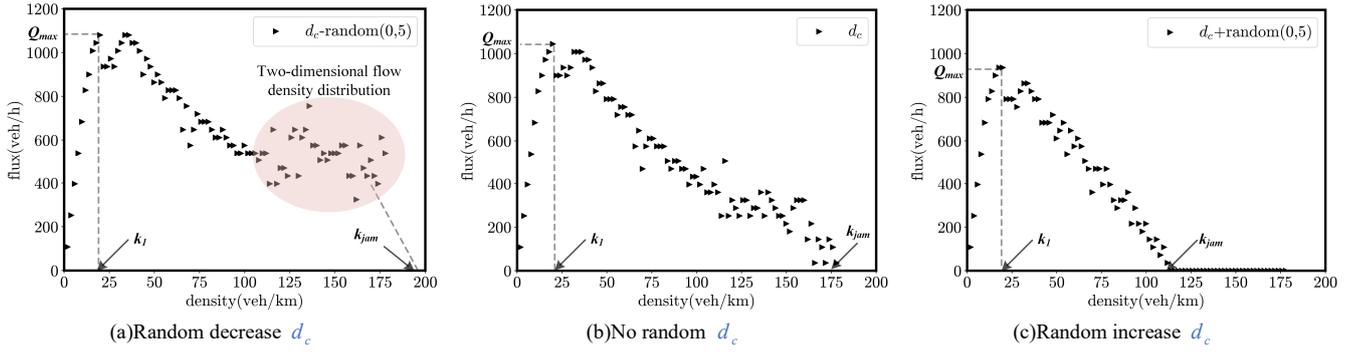

Fig. 21 Flow-density diagrams of Sigmoid-IDM with random $d_c$

As shown in **Fig. 21**, the time-varying of the cautious following distance does not affect the stable driving of the vehicle free flow, and the free flow slope is always kept at $v_0$, which is consistent with the fundamental diagram drawn by the steady-state equilibrium equation in Section III.C. Due to the introduction of random $d_c$, the degree of driver aggressiveness increases as $d_c$ decreases, and **Fig 21(a)** demonstrates significant inter- and intra-driver heterogeneity, presenting a two-dimensional flow density state when the density is larger. As shown in **Fig. 21(c)**, as $d_c$ increases, driver caution increases to initiate a slower response, the critical density $k_1$ in the flow density diagram, the blockage density $k_{jam}$ moves left, and the maximum capacity $Q_{max}$ decreases.

*E. Virtual Start-stop Scene Co-simulation*

Carsim delivers the accurate, detailed and efficient methods for simulating the performance of vehicles. To verify the traceability of the planned path by Sigmoid-IDM, a joint simulation is conducted based on MATLAB/Simulink and Carsim[65],[66]. It is assumed that the leading vehicle always remains stationary. Initial velocity and acceleration of the other vehicles are all zero. The length of the road is 100 meters and 6 vehicles are evenly distributed on the road with 14m interval. The road friction coefficient is set to 0.85, and the vehicle length is set to 2.4 m. The closed-loop velocity controller is used to control vehicle When the simulation starts, all vehicles start at the same time. Acceleration is generated by sigmoid-IDM respectively.

The visualization of platoon simulation is shown in **Figs.22(a)and(b)**. The trajectory, velocity, acceleration time and mean absolute control error per second of platoon are shown in **Figs.22(c), (d)and(e)**. As shown in TABLE III, the absolute values of the average control errors of the velocity, displacement and acceleration of the fleet are small, which proves that the trajectory of Sigmoid-IDM simulation can be applied to the actual stop-and-go scene.

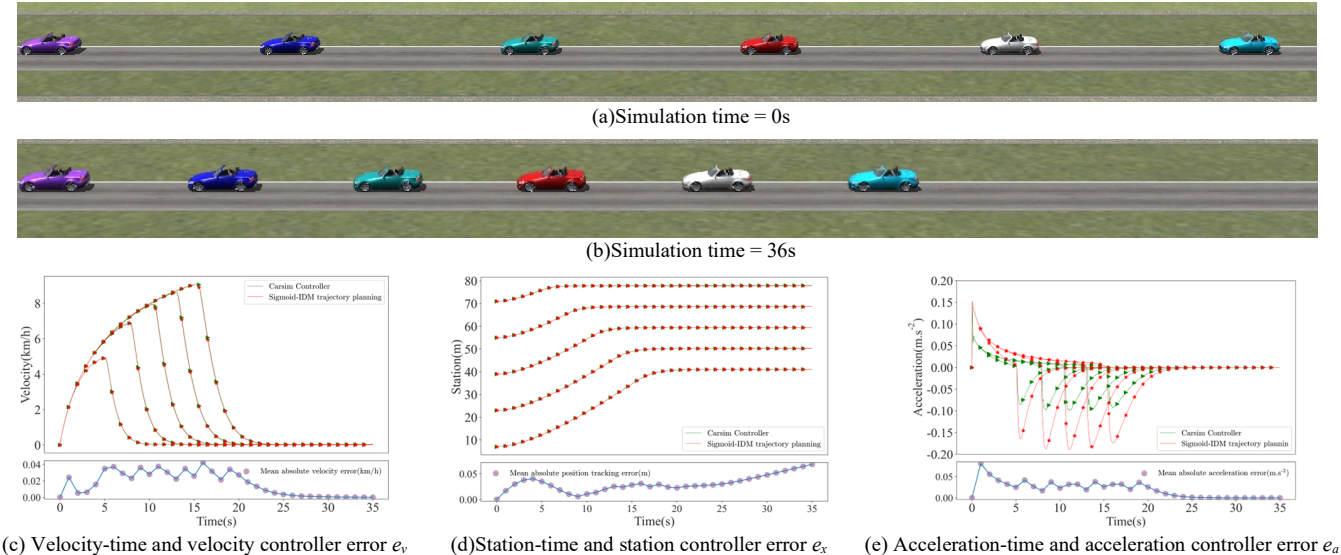

Fig. 22. Simulation of platoon Sigmoid-IDM with Carsim

**TABLE III**
**FLEET AVERAGE ABSOLUTE ERROR**

| Mean absolute error (The whole simulation) | No.2 | No.3 | No.4 | No.5 | No.6 |
|---|---|---|---|---|---|
| Mean absolute position tracking error $e_x$ | 0.014 | 0.032 | 0.015 | 0.083 | 0.021 |
| Mean absolute acceleration error $e_a$ | 0.026 | 0.026 | 0.021 | 0.021 | 0.002 |
| Mean absolute velocity error $e_v$ | 0.025 | 0.023 | 0.020 | 0.016 | 0.011 |

## VI. CONCLUSION

This paper introduces the Sigmoid-IDM, a CF model that incorporates a Sigmoid function and cautious driving factors. The Sigmoid function is used to enhance the start-following characteristics of the model, while the cautious driving factors improve the output strategy of the spacing term and stabilize the steady-state velocity in free flow. In the stability analysis, this paper analyzes the rationality of the following model and the local stability, and analyzes the string stability characteristics under different parameters according to the string stability criterion. The model has the feature of adjusting the degree of driver caution by changing the model parameters, which can better describe the heterogeneity among drivers and within drivers. In addition, the model uses Hefei data and High D data to simulate three common traffic flow scenarios: free-flow scenario, start-up scenario, and oscillation scenario, and is calibrated by minimizing spacing errors. Specifically, compared with IDM, Sigmoid-IDM reduces spacing RMSE by 46.71% and 30.48% in start-up scenario and oscillation scenario respectively, reduces fuel consumption error and comfort error by 60.1% and 49.1% respectively. Additionally, the RMSE for velocity in the free flow is reduced by 27.1%. After analysis, Sigmoid-IDM improves the trajectory anthropomorphism by improving the cautious acceleration characteristics and the asymmetric driving characteristics of IDM. In simulation, the ability of the Sigmoid-IDM to reproduce the traffic oscillations as well as the fundamental diagram was verified by the circular road simulation. The traceability of the model planning trajectory was verified by Simulink-Carsim joint simulation.

In terms of future work, it would be interesting to further explore the potential applications of the Sigmoid-IDM in various traffic scenarios and to investigate its performance in comparison to other CF models, especially the current dynamic random parameters of the CF models such as 2D-IDM. Additionally, further research could be conducted to optimize the model parameters for different driving styles and traffic conditions.

## APPENDIX A

Two brief simulations verify the IDM over-acceleration and negative velocity limitations in Section II.A. The IDM with parameters $a = 3$; $b = 2$; $v_0 = 10$; $T = 1.6$; $s_0 = 5$; Both the preceding vehicle and the following vehicle initially remain stationary, and the distance between the two vehicles is D.

**Negative velocity.** We will show that the IDM can provide negative velocity to the follower, even though the preceding vehicle may be traveling at positive velocity.

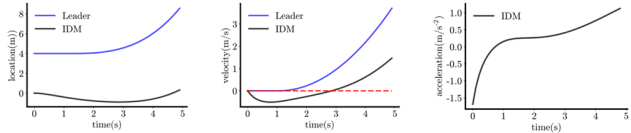

**(A. Fig 1) Left**: vehicles' positions, **middle**: vehicles' velocities and **right**: follower's acceleration. (D = 4)

**Excessive acceleration.** We will show that IDM can provide over-acceleration to a following vehicle, despite the small distance between the vehicle preceding and the vehicle following.

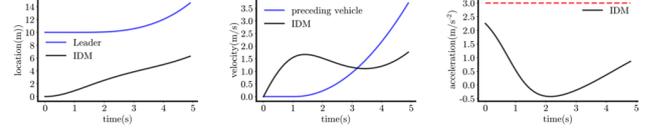

**(C. Fig 2) Left**: vehicles' positions, **middle**: vehicles' velocities and **right**: follower's acceleration. (D = 10)

**Analysis of Sigmoid-IDM to avoid negative velocity**

After the analysis above, negative speed tends to occur when the initial vehicle spacing $S(0)$ is less than the safe spacing $s_0$. When the vehicle starts, $v_n(0) = 0$, at which point the Sigmoid-IDM start equation is as follows:

$$a_n(0) = a\left(1 - \left[1 + exp\left(\lambda(S(0) - s_0 - d_c)\right)\right]^{-1}\right) \quad (A.1)$$

Since $S(0) < s_0$ at this point, (A.1) can be reduced to

$$a_n(0) = a\left(1 - \left[1 + exp\left(\lambda(-d_c + \Delta)\right)\right]^{-1}\right) \text{ and } \Delta < 0 \quad (A.2)$$

Because $\left[1 + exp\left(\lambda(-d_c + \Delta)\right)\right]^{-1} \to 1$, (A.2) can be reduced to

$$a_n(0) \to 0 \quad (A.3)$$

At this point the initial acceleration is very small can be regarded as the vehicle does not start.

## APPENDIX B

In APPENDIX B, we give the results of fitting 15 phased vehicles in Section III using the Sigmoid function in MATLAB's curve fitting library along with the parameters.

TABLE IV

$f^\lambda(x) = a\left[1 + exp\left(-\lambda(x - x_0)\right)\right]^{-1}$ for starting acceleration-spacing fitting parameters and effects.

| No. | $a$ | $\lambda$ | $x_0$ | R | AR | RMSE |
|-----|-----|-----------|-------|-----|-----|------|
| 2 | 0.06707 | 1.956 | 58.69 | 0.963 | 0.961 | 0.175 |
| 3 | 6.261 | 1.978 | 18.27 | 0.974 | 0.973 | 0.146 |
| 4 | 0.6563 | 0.6248 | 8594 | 0.939 | 0.937 | 0.057 |
| 5 | 0.4356 | 11.74 | 10.11 | 0.967 | 0.966 | 0.037 |
| 6 | 1.366 | 6.293 | 20.58 | 0.785 | 0.778 | 0.218 |
| 8 | 1.7 | 50.25 | 2.229 | 0.907 | 0.906 | 0.037 |
| 9 | 1.867 | 17.79 | 0.04534 | 0.994 | 0.994 | 0.062 |
| 11 | 1 | 44.14 | 0.116 | 0.975 | 0.974 | 0.060 |
| 12 | 0.8508 | 8.026 | 2.55 | 0.999 | 0.999 | 0.014 |
| 13 | 2.086 | 12.37 | 13.88 | 0.990 | 0.989 | 0.007 |
| 14 | 1.283 | 13.35 | 2.901 | 0.995 | 0.995 | 0.039 |
| 17 | 1.107 | 40.88 | 9.351 | 0.806 | 0.800 | 0.216 |
| 23 | 1.2 | 113.7 | 0.2399 | 0.920 | 0.919 | 0.142 |
| 24 | 1.3 | 33.44 | 0.8585 | 0.903 | 0.902 | 0.146 |
| 25 | 0.5448 | 6.568 | 18.93 | 0.838 | 0.833 | 0.077 |

## APPENDIX C

Since the Sigmoid-IDM expression is a piecewise function, it is necessary to prove its continuity. $f(t) = a_n(t)$ is a piecewise function in model. If $a_n(t)$ has the first kind of discontinuity point or continuity, it can be proved that $\int_0^t a_n(t)dt$

is continuous. $v_n(t)$ is also continuous according to $v_n(t) = \int_0^t a_n(t)dt$. Assume that

$$f_1(t) = a[1-(\frac{v_n(t)}{v_0})^4], f_2(t) = -a[\frac{S^*}{S(t)}]^2, f_3(t) = -a\left[\frac{1}{1+e^{\lambda S(t)-S^*-d_c}}\right].$$

$$a_n(t) = g(f_1(t), f_2(t), f_3(t)) = \begin{cases} f_1(t) + f_2(t) & S^* \geq S(t) \\ f_1(t) + f_3(t) & S^* < S(t) \end{cases} \quad (B.1)$$

$$\lim_{t \to t_0^{\pm}} g(f_1(t), f_2(t), f_3(t)) = \lim_{t \to t_0^{\pm}} f_1(t) + \lim_{t \to t_0^{\pm}} f_2(t)$$

$$= a[1-[\frac{v_n(t_0)}{v_0}]^4 - [\frac{S^*}{S(t_0)}]^2] = -a[\frac{v_n(t_0)}{v_0}]^4$$

$$\lim_{t \to t_0^{\pm}} g(f_1(t), f_2(t), f_3(t)) = \lim_{t \to t_0^{\pm}} f_1(t) + \lim_{t \to t_0^{\pm}} f_3(t)$$

$$= a[1-[\frac{v_n(t_0)}{v_0}]^4 - \lim_{t \to t_0^{\pm}} \frac{1}{1+e^{\lambda S(t)-S^*-d_c}}]$$

$$= -a(\frac{v_n(t_0)}{v_0})^4$$

$$g(f_1(t_0), f_2(t_0), f_3(t_0)) = f_1(t_0) + f_2(t_0) = -a(\frac{v_n(t_0)}{v_0})^4 \quad (B.2)$$

From the above conditions:

$$\lim_{t \to t_0^{-}} g(f_1(t), f_2(t), f_3(t)) = \lim_{t \to t_0^{+}} g(f_1(t), f_2(t), f_3(t))$$

$$= -a[\frac{v_n(t_0)}{v_0}]^4 = g(f_1(t_0), f_2(t_0), f_3(t_0))$$

(B.3)

It can be proved that $a_n(t)$ is continuous at $t_0$, and then it is proved that $a_n(t)$ and $v_n(t)$ are continuous in the time by Eq. (B.3) The continuity of $v_n(t)$ ensures that the Sigmoid-IDM does not appear breakpoint in the acceleration and deceleration conversion.